\documentclass{ctr_summer}

\usepackage{ctrfont}
\usepackage{natbib}
\usepackage{undertilde}
\usepackage{color}

\usepackage{graphicx}

\usepackage{amsmath,rotating}
\usepackage{dashrule}
\usepackage{undertilde}

\usepackage{url}

\usepackage{color}

\newcommand{\dd}[2]{ \frac{d {#1}}{d {#2} } }
\newcommand{\pd}[2]{ \frac{\partial {#1}}{\partial {#2}} }

\usepackage{booktabs}

\title{Toward a chaotic adjoint for LES}

\shorttitle{Toward a chaotic adjoint for LES}

\author{P. J. Blonigan\footnote[1]{NASA Postdoctoral Program, Universities Space Research Association}, P. Fernandez\footnote[2]{Department of Aeronautics and Astronautics, MIT}, S. M. Murman\footnote[3]{NASA Ames Research Center}, Q. Wang\footnotemark[2], G. Rigas\footnote[4]{Department of Mechanical and Civil Engineering, Califonia Institute of Technology}, \and L. Magri}

\shortauthor{Blonigan et~al.}

\begin{document}

\setcounter{page}{1}

\maketitle

Adjoint-based sensitivity analysis methods are powerful tools for engineers who use flow simulations for design. However, the conventional adjoint method breaks down for scale-resolving simulations like large-eddy simulation (LES) or direct numerical simulation (DNS), which exhibit the chaotic dynamics inherent in turbulent flows. Sensitivity analysis based on least-squares shadowing (LSS) avoids the issues encountered by conventional methods, but has a high computational cost. The following report outlines a new, more computationally efficient formulation of LSS, non-intrusive LSS, and estimates its cost for several canonical flows using Lyapunov analysis. \\

\hrule

\section{Introduction}
Scale-resolving simulations such as LES are necessary for engineering design and flow analysis, most notably flows in which jets, wakes, and separation dominate. In these cases, the Reynolds-averaged Navier-Stokes (RANS) solvers typically used by today's engineers often fail to accurately capture the relevant flow physics \citep{Duchaine:2015:turbo}. At the same time, engineers are interested in gradient-based design optimization, error estimation, and uncertainty quantification with flow simulations. All of these require efficient approaches for sensitivity analysis. Unfortunately, conventional sensitivity analysis approaches such as the adjoint method do not compute accurate sensitivities for statistically stationary quantities of interest in scale-resolving turbulent flow simulations like LES or DNS \citep{Blonigan:2016:AIAA}. This is because unlike RANS, LES and DNS resolve the chaotic dynamics of turbulent fluid flows \citep{Keefe:1992:lya} and the adjoint method computes unusable sensitivities for chaotic systems \citep{Lea:2000:climate_sens}. 


The recently proposed least-squares shadowing (LSS) method has shown great promise for computing accurate sensitivities of statistically stationary quantities in chaotic dynamical systems \citep{Wang:2014:LSS2}. Most recently, LSS was successfully applied to chaotic flow around a two-dimensional airfoil by \cite{Blonigan:2016:AIAA}. This study showed that LSS can compute accurate gradients, albeit at the cost of requiring large amounts of memory and wall-clock time for a case with only around 10K degrees of freedom (DoF). 

The following report analyzes a new formulation of LSS, called non-intrusive LSS (NILSS), that seeks to reduce the computational cost of LSS \citep{Ni:2016:AIAA}. Specifically, the cost of NILSS is estimated for several flow simulations using Lyapunov analysis, as the cost of NILSS scales with the number of positive Lyapunov exponents. Section \ref{s:lyapunov} introduces Lyapunov analysis. Section~\ref{s:sa} provides an overview of the issues with chaotic sensitivity analysis and presents NILSS. Section~\ref{s:lyaAnalyses} presents Lyapunov analyses of several flow simulations. Section~\ref{s:conclusions} concludes this report. 

\section{Lyapunov analysis}
\label{s:lyapunov}

To explain Lyapunov analysis the following dynamical system is considered
\begin{equation}
\frac{du}{dt} = f(u; s), \qquad u(0) = u_0,
\label{e:dyn_sys}
\end{equation}
\noindent where $u$ is a length $n$ vector of state variables and $s$ is some system parameter. For a three-dimensional compressible flow simulation, $u$ contains the five conserved quantities at all DoF (grid points). The parameter $s$ could be a flow parameter like the freestream Mach number or a geometric parameter such as chord length. 

For the system in Eq. \eqref{e:dyn_sys}, there exist Lyapunov covariant vectors $\psi^1(u),\psi^2(u),...,\psi^n(u)$ corresponding to each Lyapunov exponent $\Lambda^i$, which satisfy the evolution equation \citep{Ginelli:2007:Lya}
\begin{equation}
\frac{d}{dt}\psi^i[u(t)] = \frac{\partial f}{\partial u}\bigg|_{u(t)} \psi^i[u(t)] - \Lambda^i \psi^i[u(t)]. 
\label{e:lya_cov_def}
\end{equation}
\noindent Note that Eq. \eqref{e:lya_cov_def} is a linearization of Eq. \eqref{e:dyn_sys} with an additional term $-\Lambda^i \psi^i(u(t))$. 

\begin{figure}
	\begin{center}
	\includegraphics[width=0.5\textwidth]{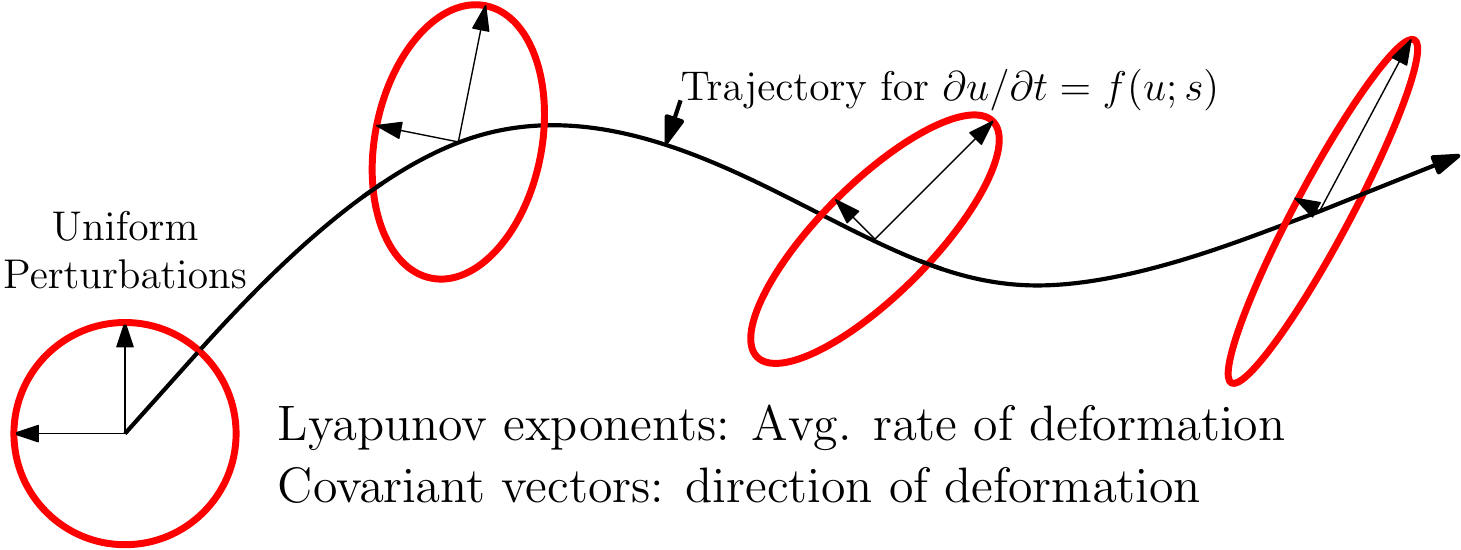}
	\caption{Schematic of Lyapunov exponents and covariant vectors in phase space. }
	\label{f:Lyapunov}
	\end{center}
\end{figure}

To understand what $\Lambda^i$ and $\psi^i$ represent, consider a sphere composed of infinitesimal perturbations $\delta u$ in phase space to the system $\frac{du}{dt} = f(u;s)$ at some time, as shown in the far left of Figure~\ref{f:Lyapunov}. As this system evolves in time, this sphere expands in some directions, contracts in some, and remains unchanged in others. The average rate at which the sphere expands or contracts is determined by the Lyapunov exponent $\Lambda^i$, and the corresponding direction of expansion or contraction is the Lyapunov covariant vector $\psi^i$. 

The magnitude and sign of the Lyapunov exponents depend on the dynamical behavior of the system being studied. If the system eventually converges to a steady state, such as any steady and laminar flow, all Lyapunov exponents are negative. This means that all perturbations to the system will decay exponentially to zero as $t \to \infty$. Different types of perturbations will decay at different rates. Specifically, an infinitesimal perturbation $\delta u(t=0) = \epsilon \psi^i(u(t=0))$ with $\epsilon <<1$ will decay at the rate $\Lambda^i$. 

An inspection of Eq.~\eqref{e:lya_cov_def} reveals that Lyapunov exponents and covariant vectors are simply the eigenvalues and eigenvectors of the linearized equation for a steady state problem (i.e., $d/dt \cdot \psi^i[u(t)] = 0$). 

For systems with a limit cycle, the Lyapunov exponents are the real part of the Floquet exponents. These systems have one Lyapunov exponent equal to zero. The zero exponent corresponds to perturbations along the cycle, or phase shifts. To see this, consider Eq. \eqref{e:lya_cov_def} with $\psi^i = f[u(t);s]$
\begin{equation}
 \frac{d}{dt}f[u(t);s] = \frac{\partial f}{\partial u}\bigg|_{u(t)} f[u(t);s] - \Lambda^i f[u(t);s].
 \label{e:Lya_zero1}
\end{equation}
\noindent By the chain rule and Eq. \eqref{e:dyn_sys}
\[
 \frac{d}{dt}f[u(t);s] = \frac{\partial f}{\partial u}\bigg|_{u(t)} \dd{u}{t} = \frac{\partial f}{\partial u}\bigg|_{u(t)} f[u(t);s]. 
\]
\noindent Therefore, Eq. \eqref{e:Lya_zero1} simplifies to $\Lambda^i = 0$. This shows that $f[u(t);s]$ is a covariant vector corresponding to $\Lambda = 0$. Since $du/dt = f[u(t);s]$, this covariant vector is tangent to the limit cycle. 

A strange attractor is similar to a limit cycle, but it has at least one positive Lyapunov exponent \citep{Ginelli:2007:Lya}. The positive exponents are responsible for the butterfly effect, a colloquial term for the large sensitivity to initial conditions exhibited by chaotic systems. A similar phenomenon occurs if a trajectory with a slightly perturbed parameter $s+\delta s$ has the same initial condition as an unperturbed trajectory with parameter $s$: the two trajectories are initially close but eventually grow apart. This results in very different instantaneous states after some time. In Figure~\ref{f:Lyapunov}, this is represented by the stretching of the sphere. The positive exponents present in chaotic systems are responsible for the issues encountered by conventional sensitivity analysis, as will be shown in Section~\ref{s:sa}. 

Lyapunov exponents and covariant vectors can be computed for numerical simulations. For the results presented in this report, Lyapunov exponents are computed using the algorithm presented in \cite{Benettin:1980:Lyapunov}. 



\section{Chaotic sensitivity analysis}
\label{s:sa}

\subsection{Breakdown of conventional sensitivity analysis}

When designing a system with unsteady flow, engineers are often interested in a time-averaged quantity $\bar{J}$,
\begin{equation}
\bar{J}(s) = \frac{1}{T} \int_{t_0}^{t_0+T} J[u(t;s);s] \ dt,
\label{e:time_avj_obj}
\end{equation}
\noindent where $J[u(t;s);s]$ is some instantaneous quantity of interest, which could be the lift or drag on an airfoil. In many cases, including applications with turbulent flow, engineers are interested in infinite time averages, $\bar{J}$ as $T \to \infty$. Since the exact evaluation of this is not computationally feasible, the infinite time average is approximated with a choice of $T$ that ensures $\bar{J}(s)$ is nearly stationary (does not vary with $T$) \citep{Oliver:2014:DNS_UQ}. 

Sensitivities with respect to the parameter $s$ can be computed using the following equation obtained by differentiating Eq.~\eqref{e:time_avj_obj}
\begin{equation}
\dd{\bar{J}}{s} = \frac{1}{T} \int_{t_0}^{t_0+T} \left( \left\langle \pd{J}{u}, v \right\rangle + \pd{J}{s} \right) \ dt, \qquad v \equiv \pd{u}{s},
\label{e:conv_fwd_grad}
\end{equation}
\noindent where $\langle \cdot, \cdot \rangle$ is the inner product and all variables on the right hand side are time dependent. 

Conventionally, the tangent solution, $v$, is obtained from the linearization of Eq. \eqref{e:dyn_sys}, referred to as the tangent equation
\begin{equation}
\dd{v}{t} = \pd{f}{u} v + \pd{f}{s}, \qquad v(t_0) = \pd{u_0}{s} = 0,
\label{e:conv_tan}
\end{equation}
\noindent The conventional approach using Eqs. \eqref{e:conv_tan} and \eqref{e:conv_fwd_grad} to compute sensitivities works for steady and periodic systems if the time horizon is an integer number of periods or if windowing is used. However, it fails for chaotic dynamical systems because chaotic systems have at least one positive Lyapunov exponent. Although the initial condition for the tangent equation (Eq. \eqref{e:conv_tan}) is zero, the term $\partial f /\partial s$ acts like a forcing term. Consider the case where $\partial f /\partial s = \delta(t-t_0) \psi^1(t_0)$, where $\delta(t)$ is the Dirac-delta function and $\psi^1(t_0)$ is the first Lyapunov covariant vector. For this case, $v(t)$ will start growing exponentially at $t=t_0$ at the rate $\Lambda^1$. In the general case, $\partial f / \partial s$ will almost always have a component in the direction of $\psi^1$, and that suffices for $v(t)$ to diverge exponentially at the rate $\Lambda^1$.


The exponential growth of $v(t)$ means that as the time horizon length $T$ is increased, the gradient computed using Eq. \eqref{e:conv_fwd_grad} grows exponentially as well. This means that conventional sensitivity analysis will compute very large, unusable sensitivities for chaotic dynamical systems like scale-resolving flow simulations including LES. This result was explained using forward sensitivity analysis, but conventional adjoint-based sensitivity analysis encounters the same exponentially growth backward in time. 

\subsection{Non-intrusive least-squares shadowing}

One approach to avoid the breakdown discussed in the previous section is LSS \citep{Wang:2014:LSS2}. LSS has been shown to compute accurate sensitivities for a number of chaotic dynamical systems, including chaotic vortex shedding from a two-dimensional airfoil \citep{Blonigan:2016:AIAA}. 

Although LSS can compute accurate sensitivities, these sensitivities come at a relatively high cost. This is because the Karush-Kuhn-Tucker (KKT) system for LSS is $nm\times mn$, where $m$ is the number of time steps and $n$ is the number of system states. For the chaotic vortex shedding flow studied by \cite{Blonigan:2016:AIAA}, this KKT system is $22.18M \times 22.18M$ for a 2,218-node mesh. Furthermore, it is difficult to solve this system, up to around 300k iterations of GMRES were required to compute sensitivities accurate to three decimal places. 

To improve the computational efficiency of LSS, we explore a reformulation of the least-squares minimization problem presented in \cite{Wang:2014:LSS2}. We use forward sensitivity analysis to present these ideas, but these ideas also apply for adjoint sensitivity analysis. 

To reduce the size of the KKT system corresponding to LSS, the following alternative minimization problem is used
\begin{equation}
\min_{v(t_i)} \sum_{i=0}^K \| v(t_i) \|_2^2, \qquad \text{s.t.} \quad \dd{v}{t} = \pd{f}{u} v + \pd{f}{s} + \eta f, \quad t \in [t_0,t_K],
\label{e:mss_min}
\end{equation}
\noindent where $\eta$ is chosen so that $\langle v(t), f[u(t);s] \rangle = 0$. Now the tangent solution $v(t)$ is minimized at $K+1$ checkpoints $t_i$ instead of at all time steps between $t_0$ and $t_K$. In this case, the minimization problem (Eq. \eqref{e:mss_min}) can be solved with a $Kn\times Kn$ KKT system. 

The size of the minimization problem can be further reduced by decomposing the tangent solution $v(t)$ into one forced and $p$ unforced components, $\hat{v}(t)$ and $V^j(t)$, respectively. 
\begin{align}
v(t) &= \hat{v}(t) + \sum_{j=1}^p \alpha^j V^j(t) \label{e:decomp}\\
\dd{\hat{v}}{t} &= \pd{f}{u} \hat{v} + \pd{f}{s} + \eta f \label{e:forced} \\
\dd{V^j}{t} &= \pd{f}{u} V^j. \label{e:unforced}
\end{align} 
This decomposition allows the minimization statement (Eq. \eqref{e:mss_min}) to be written as a minimization over the weights $\alpha^j$ at each checkpoint $t_i$ rather than the entire tangent solution $v(t_i)$. For some choices of the $V^j(t_i^+)$, a relatively small number of unforced components $p<<n$ are required and the size of the KKT system can be reduced to $Kp \times Kp$. One choice of unforced tangents $V^j(t_i^+)$ that has worked well is presented in the following NILSS algorithm outline, similar to the one consider by \cite{Ni:2016:AIAA}
\\

\begin{itemize}
\item[1.] Set $\hat{v}(t_0) =0$ and $V^j(t_0)=\text{rand}(n)$, where $\text{rand}(n)$ is a length $n$ vector of random numbers and $j=1,..,p$. Set counter $i=1$ and ensure that the $V^j$ are unitary and orthogonal to one another. 
\item[2.] Make the $V^j$ orthogonal to $du/dt$ at $t_0$ for $j=1,..,p$. 
\item[3.] Time-integrate Eqs. \eqref{e:forced} and \eqref{e:unforced} to $t=t_i$ from the initial conditions $\hat{v}(t_{i-1})$ and $V^j(t_{i-1})$, respectively. Compute and save the following integrals on the fly 
\[
g_i^j = \frac{1}{t_i-t_{i-1}}\int_{t_{i-1}}^{t_i} \pd{J}{u} V^j(t) \ dt, \qquad \hat{g}_i = \frac{1}{t_i-t_{i-1}}\int_{t_{i-1}}^{t_i} \pd{J}{u} \hat{v}(t) + \pd{J}{s} \ dt.
\]
\item[4.] Take the QR decomposition of the matrix $V_i$, a $n \times p$ matrix whose $j$th column is $V^j(t_i)$. Save $Q_i$, $R_i$, and $b_i = -Q_i^T \hat{v}(t_i)$, where $Q_i R_i = V_i$. 
\item[5.] Set $V^j(t_i)=Q_i^j$ and $\hat{v}(t_i)= \hat{v}(t_i) + Q_i b_i$, where $Q_i^j$ is the $j$th column of $Q_i$. 
\item[6.] Compute and save the scalars \[
h_i^j =  \frac{(du/dt)^T}{ (du/dt)^T du/dt}\bigg|_{t_i} V^j(t_i) , \qquad \hat{h}_i =  \frac{(du/dt)^T}{ (du/dt)^T du/dt}\bigg|_{t_i}  \hat{v}(t_i). 
\]
\item[7.] Set $V^j(t_i) = V^j(t_i) - h_i^j \dd{u}{t}|_{t_i}$ and $\hat{v}(t_i) = \hat{v}(t_i) - \hat{h}_i \dd{u}{t}|_{t_i}$.  
\item[8.] Set $i=i+1$. 
\item[9.] Repeat steps 3 through 8 until $i=K+1$. 
\item[10.] Construct and solve the following linear system from the saved matrices and vectors $R_i$ and $b_i$. 
{\small
\begin{equation}
\left[\begin{array}{ccccc|cccc}
-I & & & & & R_1^T & & & \\
& -I & & & & -I & R_2^T & & \\
& & \ddots & & & & \ddots & \ddots & \\
& & & -I & & & & -I & R_K^T\\
& & & & -I & & & & -I \\\hline
R_1 & -I & & & & & & \\
& R_2 & -I & & & & & \\
& & \ddots & \ddots & & & & \\
& & & R_K & -I & & & & 
\end{array}\right] \left[\begin{array}{c}
\alpha_1 \\
\alpha_2 \\
\vdots\\
\alpha_{K-1}\\
\alpha_K\\\hline
\beta_1 \\
\beta_2 \\
\vdots \\
\beta_K
\end{array} \right] = \left[\begin{array}{c}
0 \\
0 \\
\vdots\\
0 \\
0\\\hline
b_1 \\
b_2 \\
\vdots \\
b_K
\end{array}, \right]
\label{e:MILSS_KKT}
\end{equation}}
\noindent where $\alpha_i = [\alpha^1_i,\alpha^2_i,...,\alpha^p_i]^T$ is a length $p$ vector. Note that the Schur complement of Eq. \eqref{e:MILSS_KKT}, which is $Kp \times Kp$ instead of $2Kp \times 2Kp$, could be solved instead. 
\item[11.] Compute sensitivities from $\alpha^j_i$ using the expression
\begin{equation}
\dd{\bar{J}}{s} = \frac1K \sum_{i=1}^K \left(\hat{g}_i + \Delta J_i \hat{h}_i  + \sum_{j=1}^p (g_i^j + \Delta J_i h_i^j) \alpha_i^j \right),
\end{equation}
\noindent where $\Delta J_i = \bar{J} - J(t_i)$. 
\end{itemize}

Note that the above algorithm is similar in structure to the algorithm presented in \cite{Benettin:1980:Lyapunov} to compute Lyapunov exponents. In fact, if steps 2 and 7 are omitted, the $R_i$ matrices computed can be used to compute Lyapunov exponents as in \cite{Benettin:1980:Lyapunov},
\[
\Lambda^j = \frac{1}{T} \sum_{i=1}^K \log |[R_i]_{jj}|.
\]
Lyapunov analysis can be used to explain how the number of unforced tangents $p$ can be much smaller than the number of states $n$. To minimize Eq. \eqref{e:mss_min}, exponentially growing components of $v(t)$ must be made negligible. The number of growing components is equal to the dimension of the unstable subspace. This subspace is spanned by the Lyapunov covariant vectors corresponding to positive exponents. Therefore, to eliminate exponential growth, the columns of $V_i$ must span the unstable subspace at $t_i$. If this is accomplished, then $\alpha_i$ can be chosen such that the solutions of Eq. \eqref{e:unforced} can cancel out the exponential growth in the solution $\hat{v}(t)$ of  Eq. \eqref{e:forced}, resulting in no exponential growth in $v(t)$. Therefore, the number of unforced tangents $p$ should be at least the number of positive Lyapunov exponents, $n_+$. Past studies, including those by \cite{Pulliam:1992:chaosTransition}, \cite{Keefe:1992:lya}, and \cite{Sirovich:1991:lya}, have found $n_+$ to be a small fraction of $n$ for a range of different flows. 

To form the NILSS KKT linear system in Eq. \eqref{e:MILSS_KKT}, at least $n_++1$ tangent solutions are required over the time horizon of interest, along with $K$ QR-decompositions of a $n\times n_+$ matrix. Assuming that solving the KKT system and computing the QR decompositions are negligible in cost compared to steps 1 through 9, the cost of NILSS scales with the number of positive Lyapunov exponents $n_+$. Therefore, the cost of NILSS for a given flow simulation can be estimated by determining how many positive exponents there are. 

\section{Lyapunov analyses for flow simulations}
\label{s:lyaAnalyses}

The following sections present Lyapunov exponent spectra and some Lyapunov covariant vectors for several flow simulations. The number of positive exponents shows approximately how expensive NILSS will be for these flow simulations. 

Since not all of the flow simulations considered have linearized flow solvers, the tangent equation was approximated by a finite difference.

\subsection{Two-dimensional flow around a NACA 0012 airfoil at a high angle of attack}
\label{s:naca0012}

The two-dimensional turbulent flow around a NACA 0012 airfoil is considered first. The flow is at Reynolds number $Re_c = 2400$, freestream Mach number $M_{\infty} = 0.2$, and the angle of attack is $\alpha = 20 \ \textnormal{deg.}$, where $c$ denotes the airfoil chord, and $a_{\infty}$ is the freestream speed of sound. While two-dimensional turbulence is fundamentally different from three-dimensional turbulence, this case has been chosen due to the extensive analysis of the chaotic behavior of this flow in \cite{Pulliam:1992:chaosTransition}.

The high-order hybridized discontinuous Galerkin solver by \cite{Fernandez:2016:AIAA} is used with a 4th-order discretization in space and 3rd-order discretization in time. The computational domain is partitioned using isoparametric triangular elements, and the outer boundary is located 10 chords away from the airfoil. A run-up time of 10,000 nondimensional time units $t^{*} = t \cdot a_{\infty}/c=0.05$ is used first to drive the system to the attractor, and the Lyapunov exponent algorithm by \cite{Benettin:1980:Lyapunov} is then applied for 2,000 time units. Also, the time step is $\Delta t^{*} = 0.05$ and the time segment length is $t^* = 1.0$.

\begin{figure}
\centering
\includegraphics[width=0.56\textwidth]{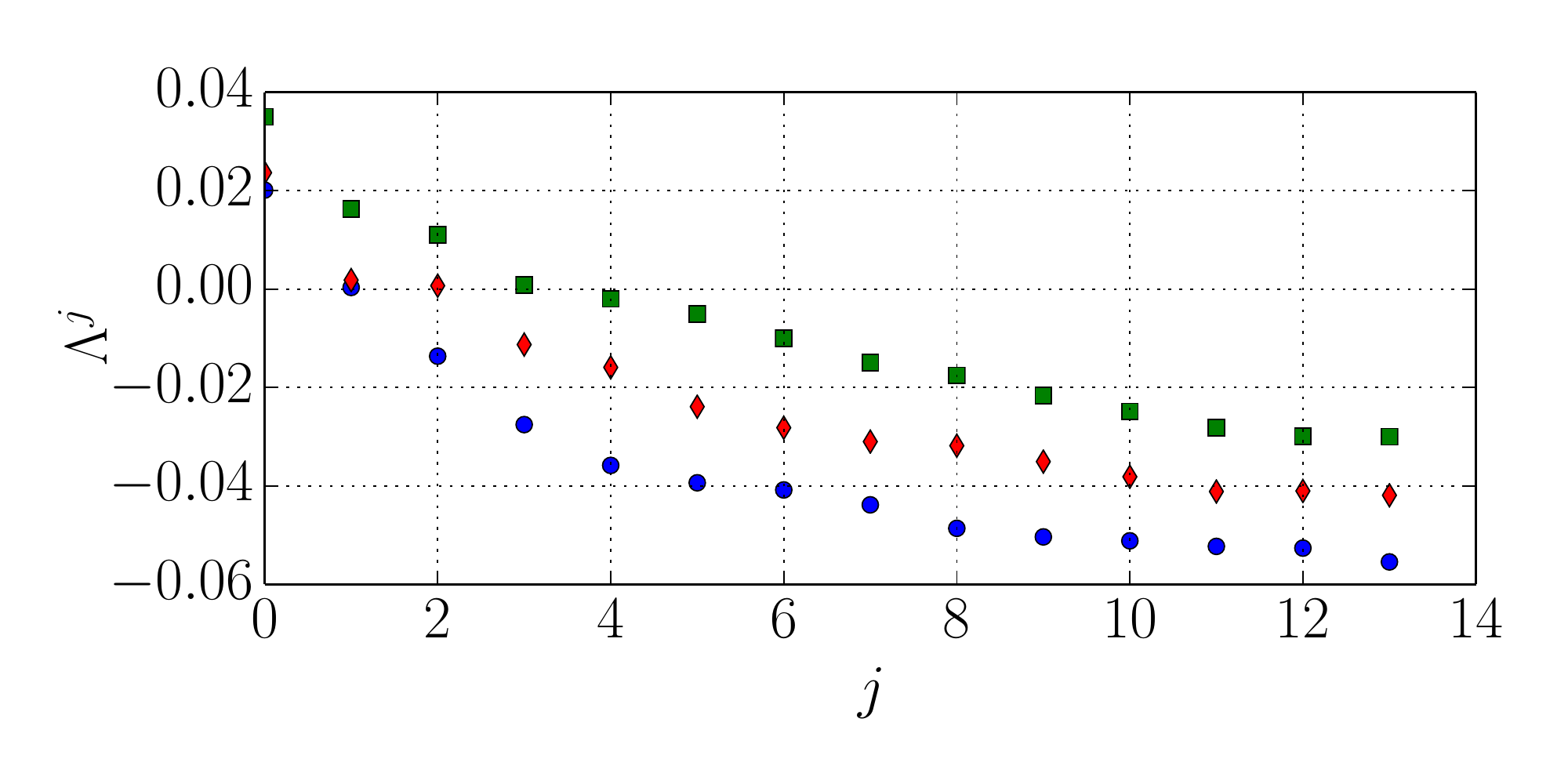}
\hspace{0.05in}
\includegraphics[width=0.3\textwidth]{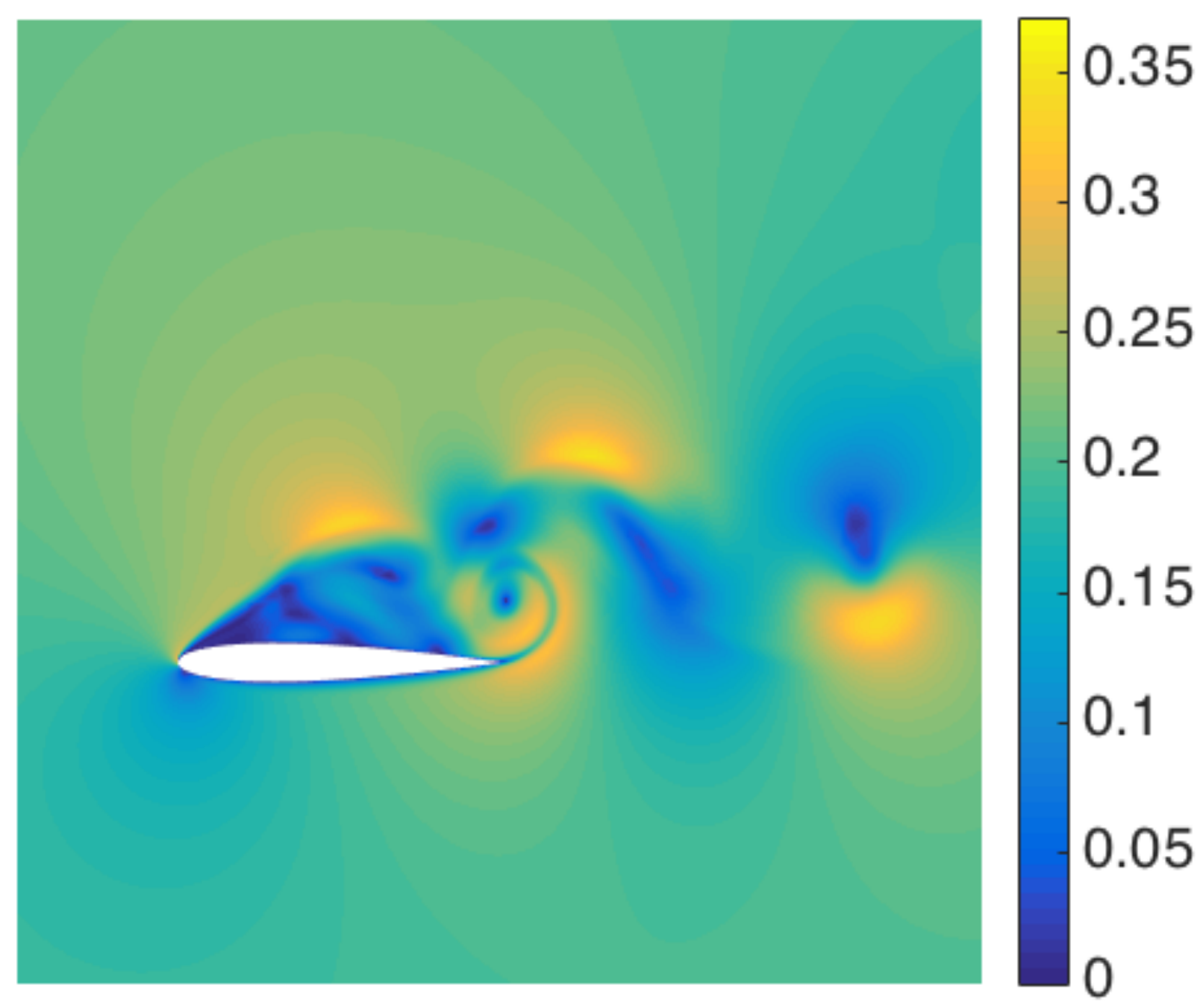}
\caption{LEFT: Lyapunov exponent spectra for the NACA 0012 airfoil coarse (blue circles), medium (red diamonds), and fine meshes (green squares). RIGHT: Snapshot of the Mach number field for the NACA 0012 airfoil.}
\label{f:NACA0012}
\end{figure}

The Lyapunov spectrum is computed on three successively finer meshes to investigate the effect of the numerical resolution on the number and magnitude of positive Lyapunov exponents. These meshes correspond to uniform refinement (x2) in both spatial dimensions, and consist of 71,680 (coarse mesh), 286,720 (medium mesh), and 1,146,880 (fine mesh) DoFs. The left panel of figure \ref{f:NACA0012} shows the 14 leading Lyapunov exponents for the three meshes considered. The right panel shows a snapshot of the Mach number field.

First, as the grid is refined, both the number and magnitude of positive Lyapunov exponents increase. This trend is also observed for the turbulent channel flow in Section \ref{s:channel}. Second, a zero Lyapunov exponent seems to be present in all discretizations. This is consistent with theoretical results and corresponds to perturbations in the $\partial f / \partial u | _{u(t)}$ direction, as discussed in Section \ref{s:lyapunov} for a limit cycle. Finally, the value of the leading Lyapunov exponent significantly differs from that in \cite{Pulliam:1992:chaosTransition}. \cite{Pulliam:1992:chaosTransition} postulated that the value of the Lyapunov exponents may largely depend on the numerical scheme used, and this seems to reinforce that observation. 


%



\subsection{Wake of an axisymmetric bluff body}
\label{s:wake}

The geometry employed is an axisymmetric bluff body with a blunt trailing edge. The length-to-diameter ratio, $L/D$, is 6.48 and the nose employs a modified super-ellipse profile with an aspect ratio of 2.5 \citep{RigasJFMr2015}. Incompressible simulations were performed with the low-Mach solver Vida (Cascade Technologies, Inc.), an edge-based unstructured finite volume solver (see \cite{RigasCTR2016} for more details). Results for a fine mesh consisting of 5M elements are presented hereafter. 

\begin{figure}
\centering
\includegraphics[width=0.3\textwidth]{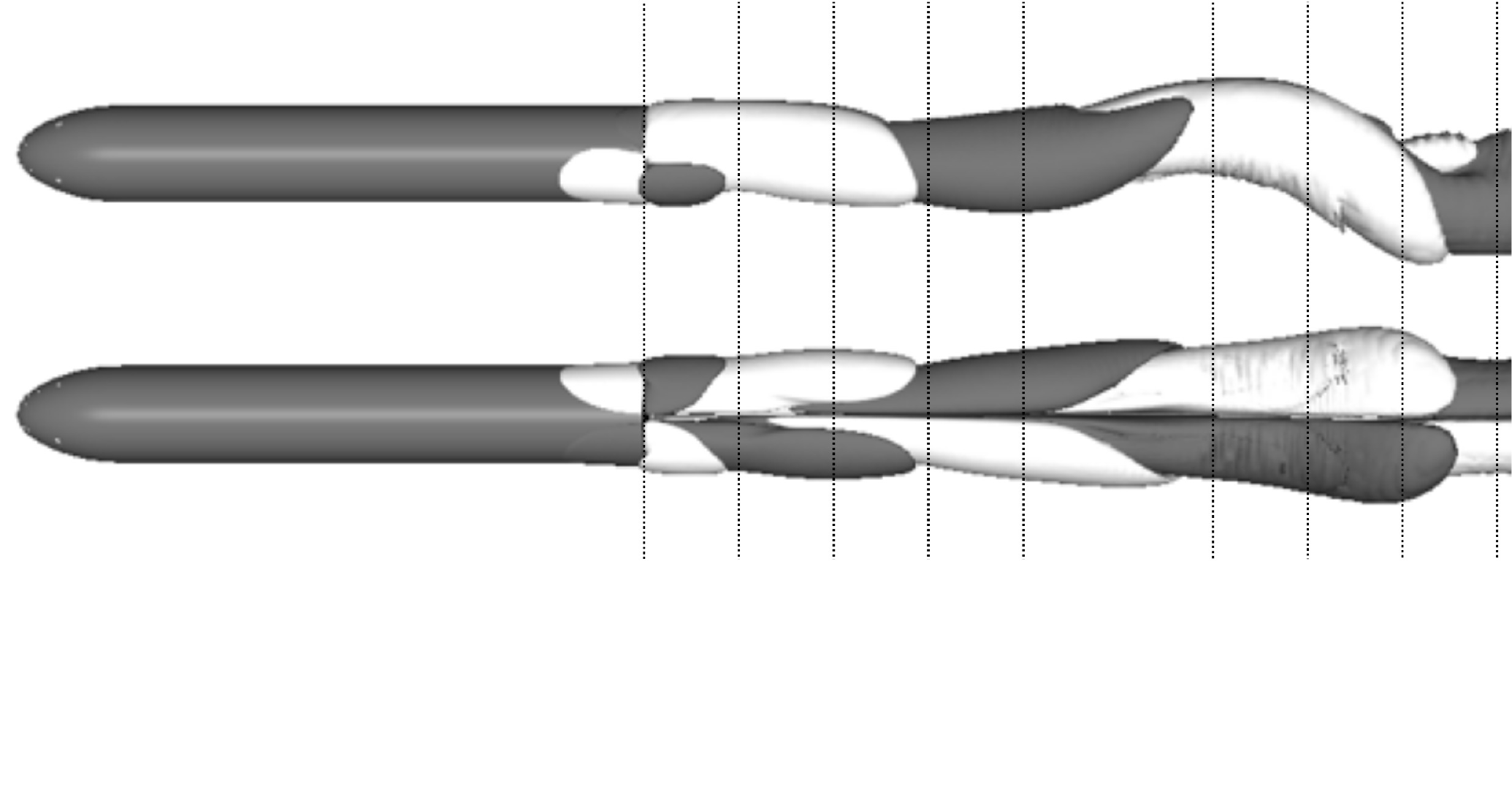}
\hspace{0.05in}
\includegraphics[width=0.65\textwidth]{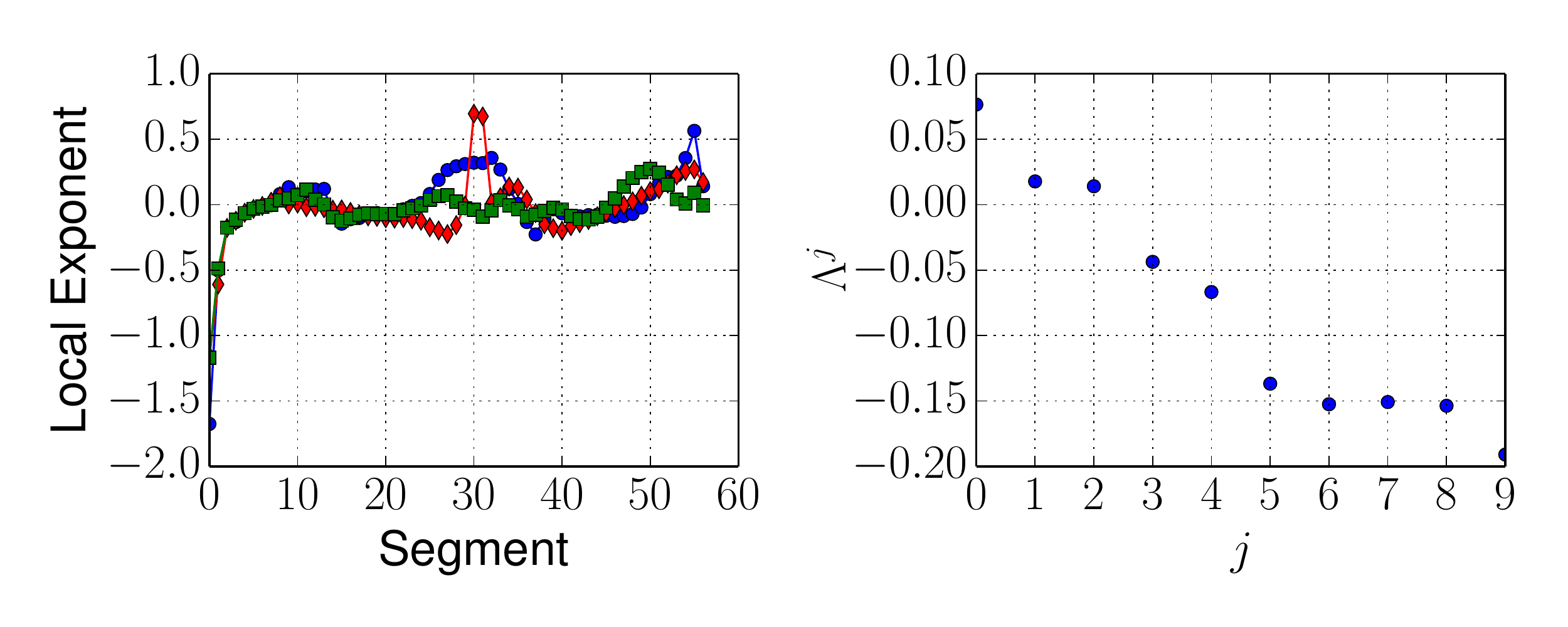}
\caption{LEFT: Streamwise vorticity for $Re=900$. Isocontours of $\omega^+= \pm0.03$ are shown up to 10D downstream of the base for top and side views. CENTER: finite-time Lyapunov exponents $j=0$ (blue circles), $j=1$ (red diamonds), and $j=2$ (green squares) for $Re=900$. RIGHT: time-averaged Lyapunov exponents, computed by averaging the finite time exponents from segment 20 to segment 55. }
\label{f:lyapunov_re900}
\end{figure}

En route to chaos, the axisymmetric bluff body wake undergoes a finite number of spatiotemporal symmetry breaking bifurcations, as shown in \cite{RigasCTR2016}. For higher values of $Re$, such as $Re_D=U_{\infty}D/ \nu_{\infty}=900$ shown on the left in Figure~\ref{f:lyapunov_re900}, the wake becomes chaotic. At $Re_D=900$ the dynamic behavior of the near wake is dominated by chaotic shedding of streamwise hairpin vortices, at a Strouhal frequency $St=0.13$. A low-frequency energetic region is identified also at $St\approx0.02$, and corresponds to irregular bursts of vorticity, indicative of the chaotic behavior of the near wake at this regime. These irregular bursts, occurring approximately every five vortex shedding cycles, manifest in Figure~\ref{f:lyapunov_re900} as isolated peaks in finite-time Lyapunov exponents with positive values. Figure~\ref{f:lyapunov_re900} also shows there are at least two positive exponents, so NILSS requires at least three forward simulations for this flow. 


\subsection{Turbulent channel flow}
\label{s:channel}

\begin{figure}
\centering
\includegraphics[width=0.46\textwidth]{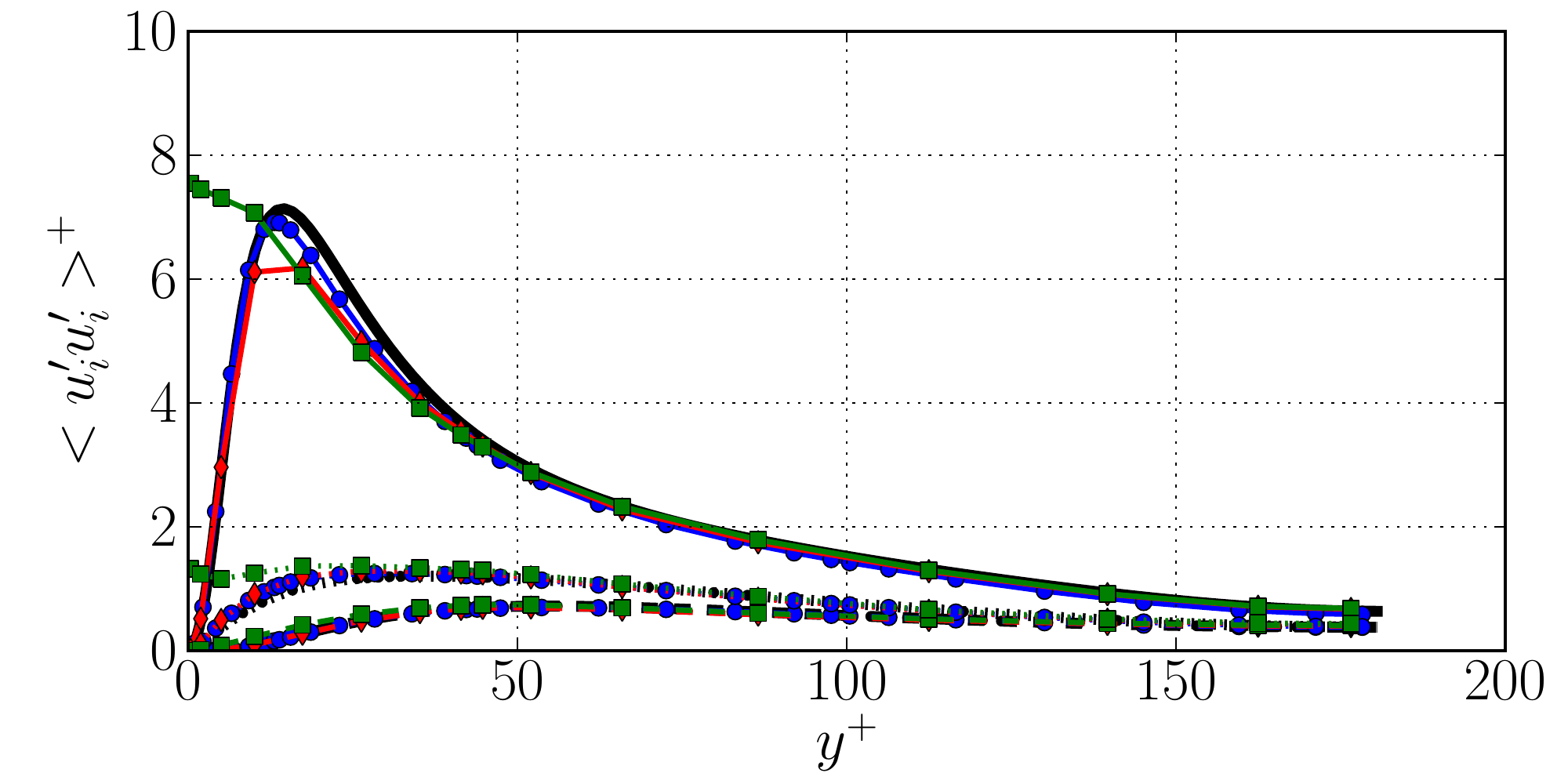}
\hspace*{0.05in}
\includegraphics[width=0.46\textwidth]{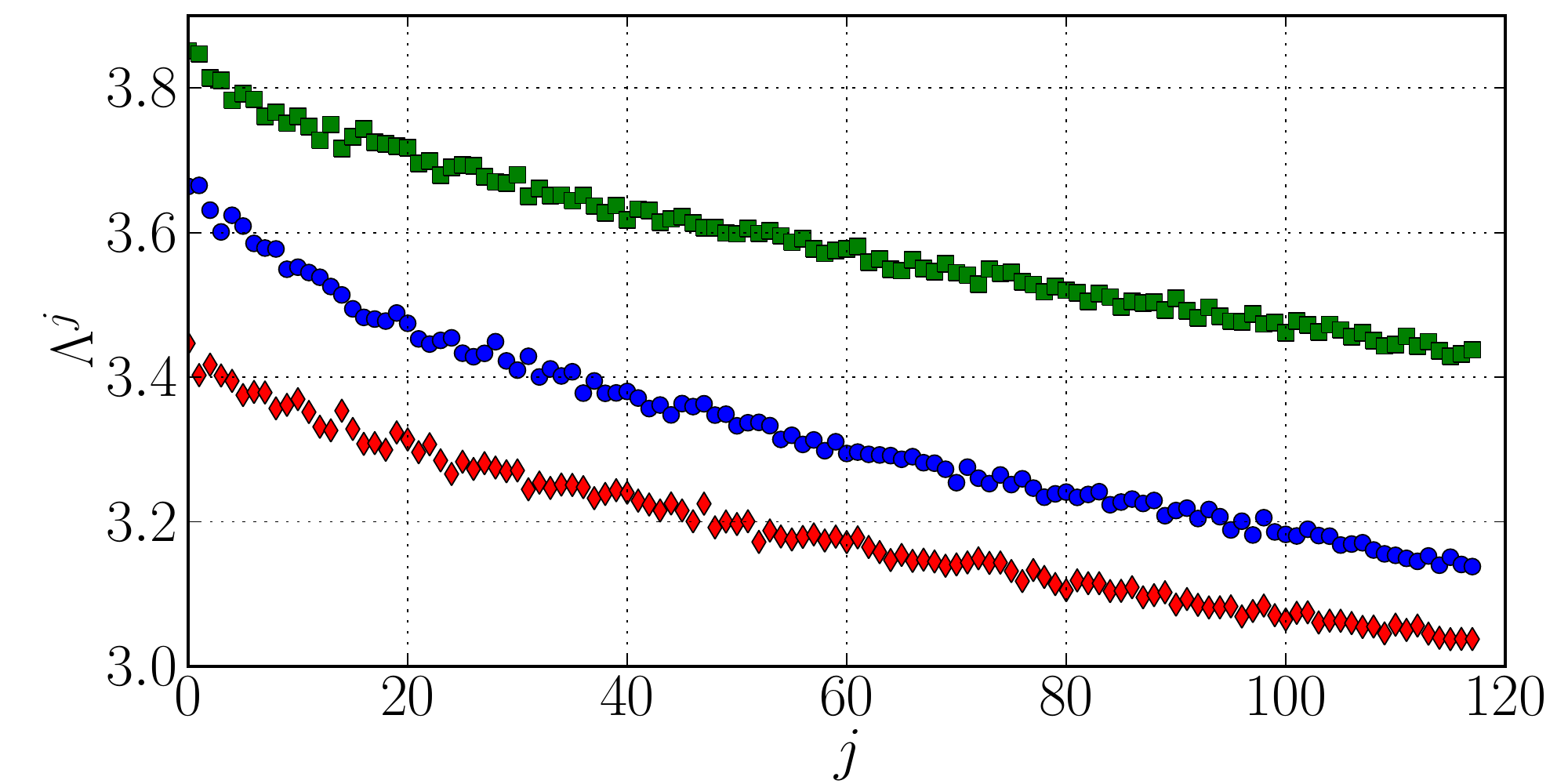}
\caption{LEFT: Velocity fluctuation profiles for all three channel flows compared with the results of \cite{Lee:2015:dns_channel} (solid black line). $u$, $v$, and $w$ fluctuations are shown by the solid, dashed, and dotted lines, respectively. RIGHT: Lyapunov exponent spectra for DNS with $\Delta y^+=2.0$ (blue circles), DNS with $\Delta y^+=5.0$ (red diamonds), and wall-modeled (green squares). All three spectra were computed with 200 time segments of length $t^+=0.273$. }
\label{f:channel}
\end{figure}

The third case considered is a turbulent channel flow with $Re_{\tau}=180$, where $Re_{\tau} = \frac{u_{\tau}\delta}{\nu}$ is the Reynolds number based on friction velocity $u_{\tau} = \sqrt{\tau_w/\rho}$ and $\delta$ is the channel half-width. The channel flow is simulated with the same space-time discontinuous Galerkin spectral-element solver used in \cite{Diosady:2014:AIAA_DNS}, with elements that are 8th order in space and 4th order in time. The domain size considered is $4\pi\delta \times 2\delta \times 2\pi\delta$ in the streamwise, wall-normal, and spanwise directions, respectively. All results presented in this report are for a mean-flow Mach number of 0.3 and a channel half-width of $\delta=1.0$. Results for fine and coarse grids are presented, with $96 \times 64 \times 80$ and $96 \times 32 \times 80$ DoFs in the streamwise, wall-normal, and spanwise directions, respectively. These grids correspond to an average spacing of $\Delta x^+ \approx 24$ and $\Delta z^+ \approx 14$ per DoF. The DoFs are arranged in the wall-normal direction so that the average spacing is $\Delta y^+ \approx 2$ and $\Delta y^+ \approx 5$ for the two grids. Despite the relatively large grid spacings, the velocity fluctuation profiles of both grids match the results from \cite{Lee:2015:dns_channel}  well, as shown in Figure~\ref{f:channel}. 

The coarse DNS case was also run with the equilibrium wall model of \cite{Carton:2014:eqwm}. Figure~\ref{f:channel} shows Lyapunov exponent spectra for all three cases. All 118 exponents are positive for all cases. A rough estimate of the total number of positive exponents $n_+$ for each case is obtained by assuming the spectrum will remain continuous and fitting a linear curve to the last 60 exponents of each spectrum, which appear to vary linearly, as done by \cite{Sirovich:1991:lya}. This extrapolation estimates $n_+=1200$ for the $\Delta y^+=2$ case, $n_+=1400$ exponents for the $\Delta y^+=5$ DNS case, and $n_+=1500$ for the wall-modeled case. 

The exponents in Figure~\ref{f:channel} are larger in magnitude than those computed by \cite{Keefe:1992:lya}. This is because the results presented here use a greater value of $Re_{\tau}$. A larger Reynolds number means that smaller length and time scales are present in the flow. Since Lyapunov exponents are inverse time scales, larger Reynolds numbers should lead to larger Lyapunov exponent magnitudes. 

The larger Reynolds number might also explain the why the estimated
number of exponents is larger than the number observed by \cite{Keefe:1992:lya} (1200 versus roughly 160 positive exponents in the finest simulation). The increase in the number of exponents is also likely due to the larger domain size in the present study ($4\pi\delta \times 2\delta \times 2\pi\delta$ versus $1.6\pi\delta \times 2\delta \times 1.6\pi\delta$). Even if the same Reynolds number was used, using a larger domain for the channel flow is similar to running multiple simulations because of the periodic boundary conditions. Therefore, doubling the domain size would double the number of positive exponents. 

The smaller relative magnitude of the exponents computed on the $\Delta y^+=5$ DNS grid is also consistent with the findings by \cite{Keefe:1992:lya}, but the larger overall number of positive exponents $n_+$ estimated for this case is not. This might indicate that the linear extrapolation used to estimate $n_+$ is inaccurate for this case. 
 
The fact that the exponents computed for the wall-modeled case are the largest in magnitude in Figure~\ref{f:channel} might be related to the larger velocity fluctuations near the wall. These larger fluctuations could be due to faster time scales in the near-wall layer caused by the absence of the no-slip boundary condition. Further study is needed to verify this idea. 

\section{Conclusions}
\label{s:conclusions}

Although conventional sensitivity analysis approaches fail for chaotic simulations like LES, LSS-based approaches are successful. The original LSS formulation is costly, but the recent NILSS approach looks to be more cost-effective. Overall, NILSS is well suited for studying low-Reynolds-number flows around bluff bodies, which appear to have relatively few positive Lyapunov exponents. The NACA 0012 and axisymmetric bluff body cases have $n_+ \sim \mathcal{O}(1)$, so only $\mathcal{O}(1)$ additional simulations are required to compute sensitivities with NILSS. Unfortunately, it seems that wall-bounded flows like the channel flow will be considerably more expensive, requiring at least $\mathcal{O}(1000)$ simulations even for low $Re_{\tau}$ cases like the ones presented in this study. Also, it seems that using a wall model does not decrease $n_+$; it increases it despite the lower mesh resolution. 

There are a large number of directions for future work, including computing more exponents for all three flows presented in this report, perhaps with higher grid resolution. Lyapunov analyzes of other flows of interest to engineers would also be valuable to give an idea of how expensive NILSS or other shadowing-based methods will cost. 

\subsection*{Acknowledgments}
The authors would like to thank Sanjeeb Bose and Frank Ham of Cascade Technologies for their assistance with the flow solver Vida, and Dr Lucas Esclapez for his help with setting up the bluff body DNS.  We are grateful to Corentin Carton De Wiart of NASA Ames for his assistance with setting up the channel flow case and the equilibrium wall model. The authors acknowledge use of computational resources from the Certainty cluster awarded by the National Science Foundation to CTR.
\bibliographystyle{ctr}

\end{document}